# The Small World Phenomenon and Network Analysis of ICT Startup Investment in Indonesia and Singapore

**Farid Naufal Aslam[1], Andry Alamsyah[2]**
School of Economics and Business, Telkom University
[1] mail@faridnaufal.com, [2] andrya@telkomuniversity.ac.id

**Abstract**

*The internet rapid growth stimulates the emergence of start-up companies based on information technology and telecommunication (ICT) in Indonesia and Singapore. As the number of start-ups and its investor growth, the network of its relationship become larger and complex, but on the other side feel small. Everyone in the ICT start-up investment network can be reached in short steps, led to a phenomenon called small world phenomenon, a principle that we are all connected by a short chain of relationships. We investigate the pattern of the relationship between start-up with its investor and the small world characteristics using network analysis methodology. The research is conducted by creating the ICT start-up investment network model of each country and calculate its small world network properties to see the characteristic of the networks. Then we compare and analyse the result of each network model. The result of this research is to give knowledge about the current condition of ICT start-up investment in Indonesia and Singapore. The research is beneficial for business intelligence purpose to support decision making related to ICT start-up investment.*

## 1. Introduction

Indonesia and Singapore are the countries with the highest number of ICT start-up investment deal in ASEAN region (Bischoff, 2015). In the first quarter of 2015, there are 38 investment deals in Singapore and 23 investments deals in Indonesia, the highest number in ASEAN than any other country in the region. This numbers of investment deals make the network of the relationship between ICT start-up with its investor in both countries become larger and more complex.

As the number of start-ups and its investor growth, the network of its relationship becoming larger and complex, but on the other side feel small. Everyone in the start-up investment network can be reached in short steps, led to a phenomenon called *Small World Phenomenon*, a principle that we are all connected by a short chain of relationships.

Currently, there is no study that investigates and compare the small world phenomena of ICT start-up investment in each country. The small world phenomena of ICT start-up investment can be investigated using network analysis methodology. Using network analysis, we can learn about each network model by measuring the properties of its network. Also, by doing a comparative study, we could know the differences between the networks and get insights from them.

## 2. Literature Review

*2.1. Network Analysis*

Network analysis is a branch of graph theory which aims at describing quantitative properties of networks of interconnected entities by means of mathematical tools. Any domain which can be described as a set of interconnected objects is a domain application for network analysis. Its methods and tools work on top of this abstraction, and as such, they are totally indifferent to the nature and properties of the entities involved, be they train stops in a railway network, individuals of a given social group bound by kinship relationship, or hosts in a computer network (Bellomi, 2009).

In this research, we use network analysis as an approach to calculate the ICT startup investment network properties. The network properties used in this research are the properties that can explain the characteristic of small world network. The definition of each network properties is shown in this table.

Table 1. The Small Network Properties Description

| Network Properties | Description |
| --- | --- |
| Degree Distribution | the probability distribution of the degree of a node over the whole network (Barabasi & Chandler, 2009). |
| Density | The fraction of number edges in network to the maximum edges possible (Newman, 2012) |
| Diameter | The largest distance recorded between any pair of nodes. (Barabasi & Chandler, 2009) |
| Average Path Length | The average distance between all pairs of nodes in the network. (Barabasi & Chandler, 2009) |
| Average Clustering Coefficient | the global value of tendency of the actors in network to form a cluster (Pandapotan, Alamsyah & Paryasto, 2015) |

*2.2. Bipartite Network*

Based on principle in graph theory, a bipartite network is a network whose nodes can be divided into two disjoint sets $U$ and $V$, where $U$ and $V$ are each independent set, such that every edge connects a node in $U$ to one in $V$ (Alamsyah & Peranginangin, 2015) . We use this kind of network to our topic, where the two set of nodes are sets of ICT startups and sets of investors that invested in that startups.

*2.3. Preferential Attachment*

A central ingredient of all models aiming to generate scale-free networks is preferential attachment, i.e., the assumption that the likelihood of receiving new edges increases with the node's degree (Barabasi & Chandler, 2009). Preferential attachment is a stochastic process that has been proposed to explain certain topological features characteristic of complex networks from diverse domains. The systematic investigation of preferential attachment is an important area of research in network science, not only for the theoretical matter of verifying whether this hypothesized process is operative in real-world networks, but also for the practical insights that follow from knowledge of its functional form (Pham, Sheridan & Shimodaira, 2015). Preferential attachment is one of the properties of the social network. Because of this preferential attachment, it will form the hub nodes in the network. The hubs will enable the formation of the small world network, because shortest path between nodes flows through this hub.

*2.4. Small World Network*

A social network exhibits the *Small World Phenomenon*, if any two individuals in the network are likely to be connected through a short sequence of intermediate acquaintances. This has long been the subject of anecdotal observation and folklore; often we meet a stranger and discover that we have an acquaintance in common. It has since grown into a significant area of study in the social sciences, in large part through a series of striking experiments conducted by Stanley Milgram and his co-workers in the 1960's (Kleinberg, 1999).

There are some characteristics that make a network can be classified as small world network, among of them is the network with degree distribution following power-law. In a small world network with a degree distribution following a power-law, deletion of a random node rarely causes a dramatic increase in mean-shortest path length (or a dramatic decrease in the clustering coefficient). This follows from the fact that the shortest paths between nodes flow through hubs, and if a peripheral node is deleted it is unlikely to interfere with the passage between other peripheral nodes. As the fraction of peripheral nodes in a small world network is much higher than the fraction of hubs, the probability of deleting an important node is very low (Barabasi, 1999).

Another class of networks is the "small worlds", two characteristics distinguish them from other network topologies: first, a small average path length, typical of random graphs (here "path" means shortest node-to-node path); second, a large clustering coefficient that is independent of network size. The clustering coefficient capture how many characteristics are identified in systems as diverse as social networks, in which nodes are people and edges are relationships (Iamnitchi et al, 2011).

## 3. Methodology

In this research, we describe the characteristic of ICT startup investment network in Indonesia and Singapore. We use network analysis methodology. This approach to measuring the properties of each network and compare them.

We use the startup investment data from *AngelList* website, a platform for startup and investors in the world to connect with each other. We collect the data of startup with funding history. We take the data that is recorded on *AngelList* until January 2016. The followings are several of data analysis process:

### 3.1. Data Collection

We crawl the startup data from *AngelList* using the *Application Programming Interface* (API) that available on its website. We only collect the startup data with funding history that located in Indonesia and Singapore up to date until January 2016.

The result of this process is JSON data that converted into CSV format containing some data variables such as startup name, startup location, startup category, startup founders and the name investors that invested in that startups, as shown in Figure 1.

| startup_name | category/_text | description | location/_text | founder/_text | investors/_text |
|---|---|---|---|---|---|
| PT Bilna | Baby Products | Indonesia Baby E-commerce | Jakarta | Eka Himawan; Ferry Tenka | DG Incubation; Vinnie Lauria; FUMIHIKO ISHIMARU; Justin Hall; Golde |
| Tokopedia | E-Commerce | Indonesia Largest Open Marke | Jakarta | William Tanuwijaya; Leontinus Tokopedia | SoftBank Ventures Korea; CyberAgent Ventures; East Ventures; netpri |
| bridestory | Weddings | Pinterest Style Wedding Vend | Jakarta | Kevin Mintaraga; Etienne Emile | William E Wijaya |
| IndoTrading.com | B2B | Indonesia Trading B2B Portal | Indonesia | Handy Chang | Rebright Partners; Founder Institute |
| Urbanesia | Mobile | Yelp for lifestyle | Jakarta | Batista Harahap; Selina Limman | Google; East Ventures |
| Qraved | Social Commerce | Great Restaurants for Less (Ins | Jakarta | Adrian Li; Steven Kim; Sean Liao | Dave McClure; Toivo Annus; 500 Startups; Rebright Partners |
| Imagi Visualize | Fashion | Visualizing your shopping | Jakarta | Christian Wirawan | |
| Pitch Clear | Startups | Elevator pitch become worldw | Jakarta | | Wawan B. Setyawan; Kane Miller |
| BantuMu | Recruiting | A LinkedIn alternatives for Pro | Bandung | Wawan B. Setyawan | Wawan B. Setyawan; Kane Miller; Galuh Sawitri; Wawan B. Setyawan |
| Jagad.co.id | Online Travel | Land Transport Marketplace | Indonesia | Amar Alpabet; Nur Hidayat | Founder Institute; Candra Agung Pratama; Alfi Setyadi Mochtar; Fahm |
| MBDC Media | Indonesia | Online Media Focusing on Vid | Jakarta | Aryo Sayogha; christian sugiono; Arianjie AZ | 500 Startups; Rebright Partners; Nusantara Incubation Fund; Sesa Nas |
| Bukalapak | E-Commerce | Indonesia ecommerce platfori | Indonesia | Achmad Zaky; Nugroho Herucahyono; M Fajrin Rasyid | Koichi Saito; YUJI HORIGUCHI; Zakka Fauzan Muhammad; Sigit Adinug |
| Valadoo | Online Travel | | Jakarta | Peter Goldsworthy; Adjie Sudradjat; Jaka Wiradisuria; Reza Muhammad | Irawan Ardhata; Syafrullah Djaya (Etjie); Maristella Gondokusumo; Pa |

Fig. 1. Node data structures from crawling process

Although we take the profile data of startups, we are confident that this activity doesn't violate any privacy since we only take data from *AngelList* via their public API and the data is accessible by anyone worldwide who register an access to the API.

### 3.2. Data Preprocessing

In data processing, we remove some redundant startups and investors data to get the appropriate data for network modeling. We also examined whether there are fraud startup data and remove them if they exist.

### 3.3. Startup Investment Network Model Construction

We create two network models for each country. We create the ICT startup investment network model with two types of node. The first node is the name of the startup and the second node is the names of investors who invest into the startup. We already put the data names of startup and investor list into two columns. The first column contains the name of the startup and the second column contains the name of investors of the startup. For example, if a startup named "startup A" invested by the investor named "Investor B" then they would be in the same row but different columns. The startup "startup A" will be in the first column and the investor "Investor B" will be in column two.

| startup_name | investor_name |
|---|---|
| SGRECX | YuhanHu |
| SGRECX | RichardKlatt |
| 3D Matters | yongtaikok |
| Share The Words | YanivCohen |
| WVI Consulting | Workuventure |
| Megafash | WillsonCuaca |

Fig. 2. Node data construction for network modeling

We open the CSV file contains two kinds of nodes into an application called *Gephi* to create the visualization of the networks. *Gephi* is an open-source network graph and analysis tool (Bastian, Heymann& Jacomy, 2009). We use *Yifan Hu* proportional layout for each network in order to display a comprehensive visualization.

*3.4. Startup Investment Network Properties Calculation*

We calculate the value of the network properties using *Gephi*. It provides some feature to measure the network properties. The measurement in *Gephi* is the *degree distribution, density, diameter, average path length,* and *average clustering coefficient* of each network.

*3.5. Startup Investment Network Comparison*

We analyze the value of the network properties to determine the pattern and the small world characteristics of each network. Then we compare the result to determine how small the networks are, which network is smaller, and which one has the best value of small world network.

## 4. Result & Analysis

The data characteristics and descriptions are as follows: before preprocessing, per January 2016, Indonesia has 554 startup data, while Singapore has 1271 startup data. After preprocessing, Indonesia startup has 182 data, while Singapore has 1025 data. When the networks formed, the Indonesia bipartite network of startups and investors consist of 182 nodes and 157 edges, while the Singapore bipartite network of startups and investors consist of 1025 nodes and 913 edges.

We compute the network properties and visualize the network. First, we conduct measurement of degree distribution of the networks. The comparison between the two networks shown in the following chart in Figure 3.

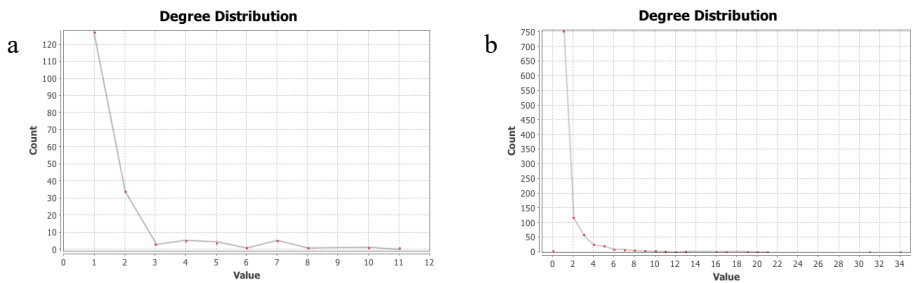

Fig. 3. (a) Indonesia ICT Startup Investment Network; (b) Singapore ICT Startup Investment Network

From the chart above, the degree distributions of the two networks follows the power-law rule, where the number of lower degree nodes is higher than the nodes with higher degree value. A small world network has a characteristic of power-law obeying degree distributions.

If we delete a random node in a small world network with power-law degree distribution, the chance of the value of average short path increases or the value of average clustering coefficient decrease is rare. This because shortest path between nodes flows through hubs. If a node around the hub is removed, it will not affect other parts of the nodes that attached to the hub. This hub node can be created because of the preferential attachment characteristic of the social network.

The comparison of the other network properties measurement is shown in the Table 2 below:

Table 2. Network Properties Comparison

| Network Properties | Indonesia | Singapore |
| --- | --- | --- |
| Density | 0.01 | 0.002 |
| Diameter | 10 | 13 |
| Average Path Length | 4.459 | 6.09 |
| Average Clustering Coefficient | 0.066 | 0.005 |

From the table 2, the ICT startup investment density network in Indonesia has the higher value than Singapore. The density of network shows us how dense the connection of every node in a network. It also can be said as the number of people who actually know each other. The higher the density value of a network means the more connections appear in every node in the network. From the two networks, Indonesia gets higher density value than Singapore. It means the possibility of the nodes to know each other in the network is higher in Indonesia, it could make the network feel smaller because most people already know each other.

Diameter measures the longest shortest path in a graph. A small world network has a low value of network diameter. Based on network comparison, Indonesia has the lower value of network diameter than Singapore.

The next property is average path length, which is the average distance between all pairs of nodes in the network (Barabasi & Chandler, 2009). The lower value of average path length indicates that it only needs short steps for a random node in the network to connect with each other. The average path length of Indonesia is 4.459, which is lower than Singapore that has 6.09 as the value of average path length.

Average clustering coefficient shows the global value of tendency of the actors in the network to form a cluster (Pandapotan, Alamsyah & Paryasto, 2015). A small world network tends to contain many clusters; this means the network should have a high clustering coefficient. From the measurement, Indonesia has the higher value of average clustering coefficient than Singapore. It shows that there is more cluster formed in Indonesia ICT startup investment network than Singapore.

The figures 2 below are the visualization of the ICT startup network investment in Indonesia and Singapore the graphs show the relationship between nodes in the network. The networks that formed are a bipartite network between startups and its investors.

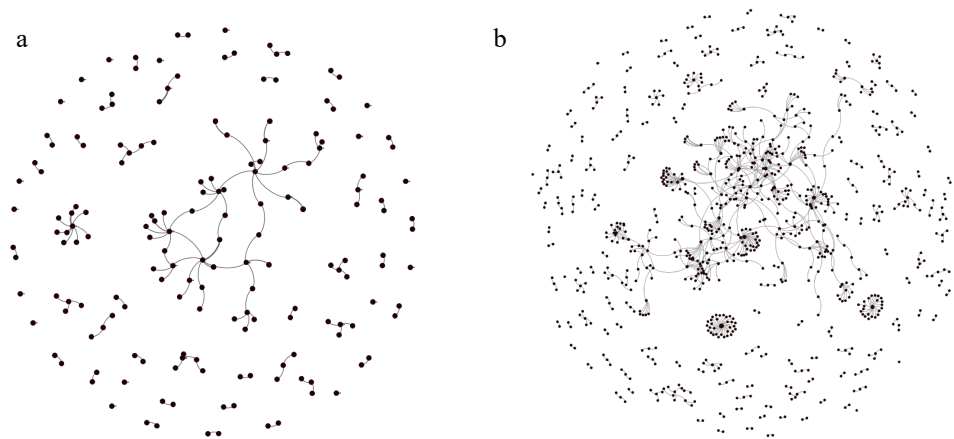

Fig. 4. (a) Indonesia ICT Startup Investment Network Visualization; (b) Singapore ICT Startup Investment Network Visualization

## 5. Conclusion & Suggestion

We have analyzed the ICT startup network investment in Indonesia and Singapore. We use network analysis to explain & compare the small world phenomena in both countries. We use degree distributions, density, diameter, average path length, and average clustering coefficient as network property measurement.

We find that ICT startup investment network in Indonesia and Singapore have characteristics of the small-world network as the network follow the common social network characteristic such as preferential attachment and power-law degree distributions. Based on the measurement, Indonesia has smaller network than Singapore. In Indonesia, the separation between each other nodes are shorter. In average, it only needs 4.459 steps to get connected with each other. The actors in Indonesia network also have a tendency to gather and eventually construct a group based on their similarity. It can be said that in Indonesia network is easier to get connected with each other than Singapore network. This also could bring the advantage for startups and investors in Indonesia to get or spread the information faster than Singapore.

This study gives an overview of the comparative study about the current ICT Startup Investment in Indonesia and Singapore using network analysis methodology. However, we can't conclude that the ICT startup investment in Indonesia is better than in Singapore because it relatively depends on many aspects. As we know the current number of startup investment deal in Singapore is higher than Indonesia, but as a network, Indonesia has more advantage of its small world characteristic than Singapore.

We can use this research to support business intelligence activity for startups or investors in each country. After we understand the model of the network, we can use this insight to decide what kind of strategy to formulate in each country related to ICT startup investment.